\documentclass[referee,sn-nature]{sn-jnl}% referee option is meant for double line spacing

\usepackage{graphicx}%
\usepackage{multirow}%
\usepackage{amsmath,amssymb,amsfonts}%
\usepackage{amsthm}%
\usepackage{mathrsfs}%
\usepackage[title]{appendix}%
\usepackage{xcolor}%
\usepackage{textcomp}%
\usepackage{manyfoot}%
\usepackage{booktabs}%
\usepackage{algorithm}%
\usepackage{algorithmicx}%
\usepackage{algpseudocode}%
\usepackage{listings}%
\usepackage{siunitx}

\usepackage{geometry}
\usepackage{nicefrac}
\theoremstyle{thmstyletwo}%

\theoremstyle{thmstylethree}%

\unnumbered% uncomment this for unnumbered level heads

\begin{document}

%\title[Bagani et al.]{Imaging strain-induced domain formation in the layered antiferromagnet CrSBr}
\title[Bagani et al.]{Imaging strain-controlled magnetic reversal in thin CrSBr }
%\title[Bagani et al.]{Imaging strain-controlled magnetic reversal in thin CrSBr}
%\title[Bagani et al.]{Imaging strain-controlled magnetic reversal in the layered antiferromagnet CrSBr}

\author[1]{\fnm{Kousik} \sur{Bagani}}
\equalcont{These authors contributed equally to this work.}

\author[1]{\fnm{Andriani} \sur{Vervelaki}}
\equalcont{These authors contributed equally to this work.}

\author[1]{\fnm{Daniel} \sur{Jetter}}

\author[2,3]{\fnm{Aravind} \sur{Devarakonda}}

\author[1]{\fnm{Märta A.} \sur{Tschudin}}
\author[1]{\fnm{Boris} \sur{Gross}}
\author[4]{\fnm{Daniel G.} \sur{Chica}}
\author[1,5]{\fnm{David A.} \sur{Broadway}}
\author[3]{\fnm{Cory R.} \sur{Dean}}
\author[4]{\fnm{Xavier} \sur{Roy}}

\author[1]{\fnm{Patrick} \sur{Maletinsky}}

\author*[1,6]{\fnm{Martino} \sur{Poggio}}\email{martino.poggio@unibas.ch}

\affil[1]{\orgdiv{Department of Physics}, \orgname{University of Basel}, \country{Switzerland}}

\affil[2]{\orgdiv{Department of Physics}, \orgname{Columbia University}, \country{USA}}

\affil[3]{\orgdiv{Department of Applied Physics and Applied Mathematics}, \orgname{Columbia University}, \country{USA}}

\affil[4]{\orgdiv{Department of Chemistry}, \orgname{Columbia University}, \country{USA}}

\affil[5]{\orgdiv{School of Science}, \orgname{RMIT University}, \country{Australia}}

\affil[6]{\orgdiv{Swiss Nanoscience Institute}, \orgname{University of Basel}, \country{Switzerland}}


\abstract{
Two-dimensional materials are extraordinarily sensitive to external stimuli, making them ideal for studying fundamental properties and for engineering devices with new functionalities. One such stimulus, strain, affects the magnetic properties of the layered magnetic semiconductor CrSBr to such a degree that it can induce a reversible antiferromagnetic-to-ferromagnetic phase transition. Given the pervasiveness of non-uniform strain in exfoliated two-dimensional magnets, it is crucial to understand its impact on their magnetic behavior. Using scanning SQUID-on-lever microscopy, we directly image the effects of spatially inhomogeneous strain on the magnetization of layered CrSBr as it is polarized by a field applied along its easy axis. The evolution of this magnetization and the formation of domains is reproduced by a micromagnetic model, which incorporates the spatially varying strain and the corresponding changes in the local interlayer exchange stiffness. The observed sensitivity to small strain gradients along with similar images of a nominally unstrained CrSBr sample suggest that unintentional strain inhomogeneity influences the magnetic behavior of exfoliated samples and must be considered in the design of future devices.}

%%================================%%
%% Sample for structured abstract %%
%%================================%%



\maketitle

\section{Introduction}\label{sec1}

The study of two-dimensional (2D) materials and their heterostructures has revealed a diverse array of electronic, optical, and magnetic properties, paving the way for groundbreaking advancements in condensed matter physics and materials science. These ultra-thin materials are exceptionally sensitive to external stimuli, such as strain, doping, and electromagnetic fields, each of which can be used to influence or control their properties. This responsiveness extends to 2D magnets, affecting parameters such as the ordering temperature, magnetic coercivity, magnetic anisotropy -- potentially even altering the magnetic easy axis or triggering magnetic phase transitions~\cite{li_tuning_2021, verzhbitskiy_controlling_2020, zhuo_manipulating_2021, siskins_magnetic_2020, wang_strainsensitive_2020}.

%Among the 2D magnets that have been studied so far, CrSBr is a promising material both for potential applications in magnetic devices and as a model system for understanding the mechanisms underlying 2D magnetism. 
CrSBr is a layered van der Waals material and an A-type antiferromagnet with a N\'eel temperature of $T_\text{N}\sim 132$~K in the bulk~\cite{goser_magnetic_1990, telford_layered_2020}. Its magnetic anisotropy is triaxial with both its intermediate and easy axes pointing in the layer plane, along the crystallographic $a$- and $b$-axis, respectively~\cite{goser_magnetic_1990}. It stands out as a direct bandgap semiconductor with strong magneto-electronic coupling~\cite{telford_layered_2020, telford_coupling_2022} and tunable magnetic properties~\cite{cenker_reversible_2022, cenker_strain-programmable_2023, telford_designing_2023,tabataba-vakili_doping-control_2024}. In addition, unlike the extensively studied Cr-trihalide family of 2D magnets~\cite{shcherbakov_raman_2018, wu_degradation_2022, mastrippolito_emerging_2021}, it displays excellent air stability~\cite{telford_layered_2020, lee_magnetic_2021, tschudin_imaging_2024}, simplifying device fabrication and experimental investigation.

The response of CrSBr to mechanical strain is particularly noteworthy: the application of strain along its $a$-axis modifies its magnetic properties, with the most significant change occurring in the interlayer exchange interaction. Specifically, compressive strain has been predicted to enhance the antiferromagnetic (AF) interlayer interaction, whereas tensile strain reduces this coupling, eventually converting it from AF to ferromagnetic (F)~\cite{bo_calculated_2023, cenker_reversible_2022, diao_strain-regulated_2023, yang_triaxial_2021, wang_origin_2023}. In fact, a controllable and reversible strain-induced AF to F phase transition has been reported under an applied uniaxial strain exceeding 1\%. This observation demonstrates the potential of CrSBr as the active element in devices, such as magnetoresistive switches or magnetic tunnel junctions, that are actuated by strain rather than applied magnetic field~\cite{cenker_reversible_2022, cenker_strain-programmable_2023}. So far, however, locally resolved measurements of how spatially varying strain or strain intrinsic to exfoliated samples influences the magnetic properties of CrSBr have not been carried out.

In this study, we directly measure the impact of strain on magnetic reversal in CrSBr flakes, demonstrating the influence of spatially varying strain. We perform nanometer-scale magnetic imaging to examine how applied magnetic fields affect the material, employing a cantilever scanning probe with an integrated superconducting quantum interference device (SQUID). We study the way in which nonuniform strain alters the magnetic hysteresis and influences the formation and subsequent evolution of magnetic domains. In order to distinguish the effects of strain from the intrinsic behavior of thin CrSBr, we compare the behavior of strained and pristine flakes with similar geometries. Furthermore, we use a micromagnetic model to simulate the magnetic evolution of a strained flake, reproducing both the spatial and field-dependent features of our measurements and gaining insight into the underlying magnetization configurations. The results show how local variations of strain can be used to create magnetic domains on demand. Moreover, they highlight the sensitivity of 2D magnetic systems to strain and how unintentional strain gradients, induced during sample fabrication or processing, can result in inhomogeneous behavior.

\section{Results} \label{sec2}

\subsection{Inducing strain via bending} \label{sec3}

Exfoliated CrSBr flakes naturally form a ribbon shape due to their anisotropic crystal structure. The long edge aligns with the crystallographic $a$-axis, the short edge with the $b$-axis, and the $c$-axis points out of the plane. This geometry is ideal for the application of in-plane strain by physically bending the ribbon. Using a recently developed method~\cite{kapfer_programming_2023}, we position a gear-shaped graphite micro-structure adjacent to one end of a CrSBr ribbon. We then slide the micro-structure over one end of the ribbon using the tip of an atomic force microscopy (AFM) cantilever, as shown in Figure~\ref{fig1}a. This action drags the end of the ribbon along with the manipulator, while the opposing end remains firmly secured to the substrate. The result is a bent ribbon that is elongated on one side and compressed on the other, as shown in Figure~\ref{fig1}b, thereby generating tensile and compressive strain, respectively, along the $a$-axis. Figure~\ref{fig1}c shows a topographic image of such a bent CrSBr ribbon with a fracture in its lower segment due to excessive bending.

To calculate the inhomogeneous strain produced in the bent CrSBr ribbon, we use a coordinate system aligned with the ribbon's crystallographic axes before bending, as shown in Figure~\ref{fig1}b. In this system, the $a$- and $b$-axis represent the original crystallographic directions and $x_a$ and $x_b$ denote positions along these directions.  We obtain a map of the strain from \( \varepsilon (x_a,x_b) = -l(x_a, x_b) ⁄ \rho(x_a) \), where $l(x_a, x_b)$ is the perpendicular distance from the neutral axis, $\rho (x_a) = (1 + (\frac{d\delta(x_a)}{dx_a})^2)^{3/2} / \frac{d^2\delta(x_a)}{dx_a^2}$ is the local radius of curvature, and $\delta(x_a)$ is the deflection of the ribbon at a given position along the neutral axis~\cite{kapfer_programming_2023}. We determine $\delta(x_a)$ by fitting the ribbon's topographic outline with a 4th-order polynomial. We then calculate a map of $\varepsilon (x_a,x_b)$, shown in Figure~\ref{fig1}d, quantifying the variation of $a$-axis strain in the sample plane. On the two long edges of the ribbon, the tensile and compressive strain reach a maximum of 0.25\%. A tiny segment of ribbon near the fracture experiences the most bending, resulting in a maximum strain of approximately 1.5\%. 

\subsection{Mapping magnetic stray field} \label{sec4}

\begin{figure}[]%
\centering
\includegraphics[width=1\textwidth]{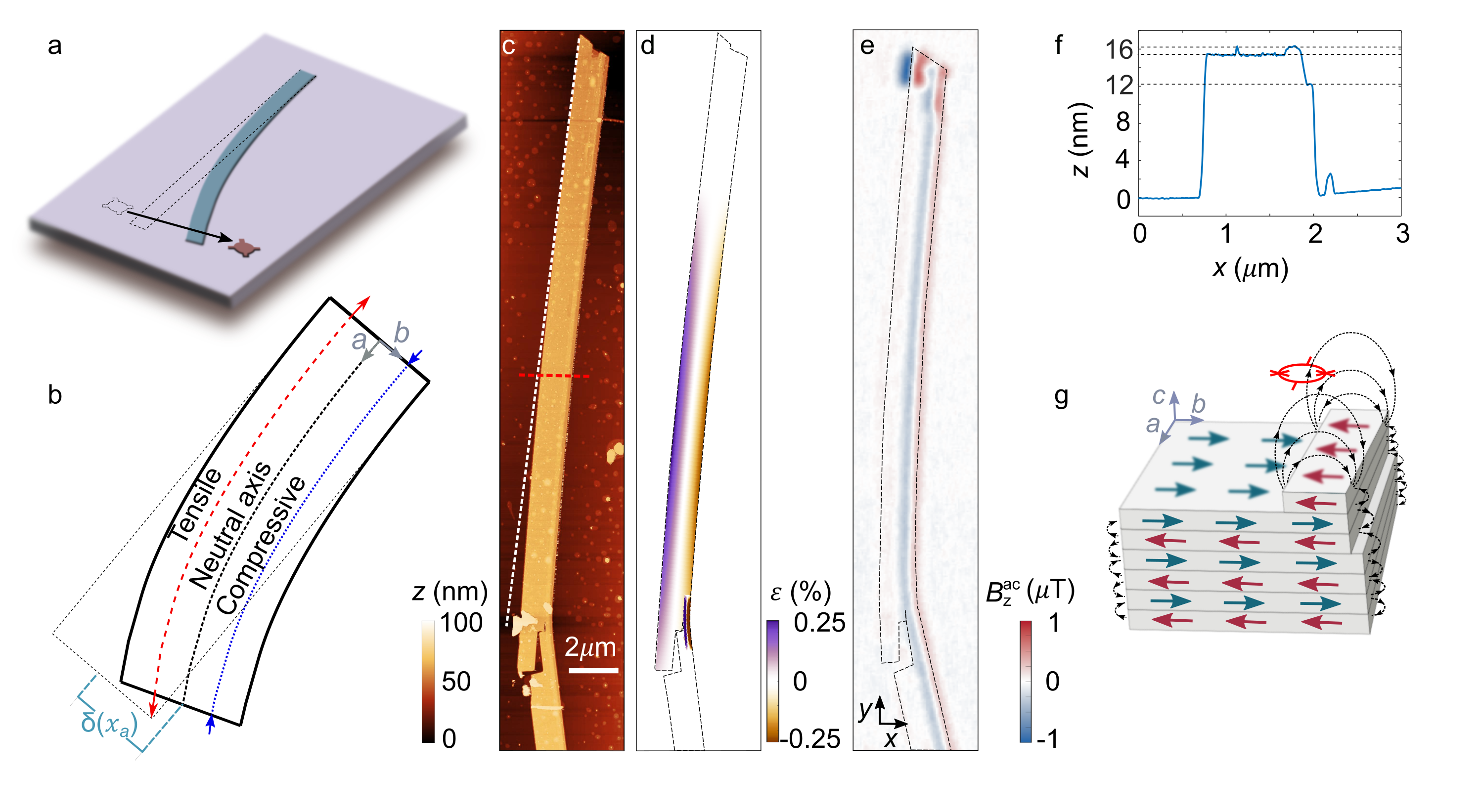}
\caption{\textbf{Strained CrSBr ribbon.} Schematic of (a) the bending procedure using a micro-manipulator and (b) the resulting strain. (c) An atomic force microscopy (AFM) image of the CrSBr ribbon after bending and (d) a map of the calculated strain induced by this bending procedure. The white dashed line in (c) represents the initial shape of the ribbon before the bending. (e) $B_z^\text{ac} (x,y)$ measured in the absence of in-plane applied field. The $x$- and $y$-axis correspond to the coordinates of the scanning probe. (f) A line-cut across the AFM image taken along the dashed red line in (c). The black dashed lines indicate the different thicknesses -- 14, 18 and 19 layers -- found within the ribbon. (g) Schematic drawing of the ribbon's AF magnetization configuration (showing fewer layers than in the actual ribbon) with magnetization shown in red and blue arrows and stray field lines as black dotted arrows . The red symbol represents the SQUID-on-lever scanning probe. The crystallographic $a$-, $b$- and $c$-axis of CrSBr are represented with grey arrows. }\label{fig1}
\end{figure}

We image the out-of-plane component of the strained ribbon's magnetic stray field $B_z$ in a plane roughly 250~nm above the sample at 4.2~K. Measurements are carried out using a SQUID-on-lever (SOL) scanning probe~\cite{wyss_magnetic_2022}, which in addition to $B_z$ allows for the measurement of $B_z^\text{ac} \propto d B_z / d z$ by demodulating the SOL response at the cantilever oscillation frequency. Due to spectral filtering, $B_z^\text{ac}$ contains less noise than $B_z$. Figure~\ref{fig1}e shows $B_z^\text{ac}(x,y)$ measured in the absence of an applied in-plane magnetic field. The image shows magnetic contrast that is primarily confined to the right edge of the ribbon.  

In the absence of an applied in-plane field, the strained ribbon should be in an AF configuration with the magnetization of each successive layer pointing along the $b$-axis and alternating its orientation. Based on previous findings~\cite{cenker_reversible_2022}, the magnitude of the applied strain is not sufficient to change this AF remanent configuration. Therefore, we expect regions with an even number of layers to produce no stray field, because the magnetization of adjacent AF layers exactly compensate, whereas regions with odd number of layers should produce a stray field equivalent to that of one uncompensated layer.

We compare our measurements to this expectation by determining the ribbon's layer thickness via AFM, as discussed in the Methods and shown in Figure~\ref{fig1}c. A line-cut of this AFM, which is plotted in Figure~\ref{fig1}f, shows that the majority of the ribbon consists of 18 layers, except for a single-layer strip 350-nm-wide running along its right edge. Immediately to the right of this strip is a 210-nm-wide region, which is 14-layers-thick. Therefore, we attribute the dipolar magnetic contrast observed near the right edge of the ribbon to the single uncompensated layer along the right edge of the ribbon and show the corresponding magnetization configuration schematically in Figure~\ref{fig1}g. Aside from this strip, the rest of the ribbon is fully compensated and produces no magnetic contrast. An exception is the region near the top end of the ribbon, which includes areas where the thickness is not uniform.

\subsection{Magnetic field dependence}\label{sec5}

\begin{figure}[]%
\centering
\includegraphics[width=1\textwidth]{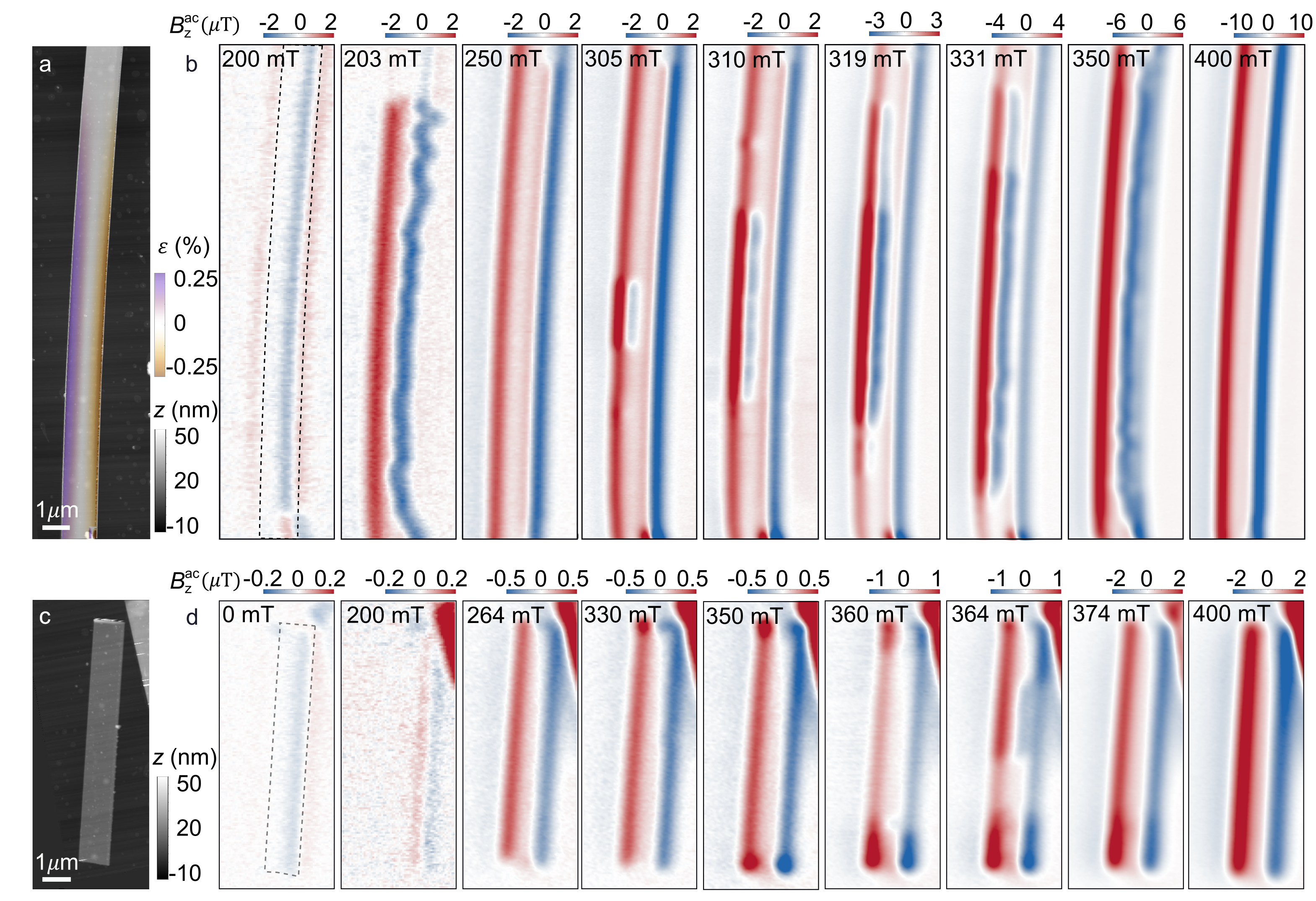}
\caption{\textbf{Magnetic field dependence of strained and pristine ribbons.} (a) A zoomed-in AFM image of the strained ribbon presented in Figure~\ref{fig1}c overlaid with a map of the calculated strain. (b) $B_z^\text{ac}(x, y)$ measured over this region under increasing  in-plane applied field $\mu_0H_x$ as indicated in the top left of each image. Dashed outlines in the first field map denote the physical boundaries of the ribbon.  (c) AFM image of the pristine ribbon and (d) the corresponding $B_z^\text{ac}(x, y)$ measured under increasing $\mu_0H_x$.}\label{fig2}
\end{figure}

We next image the stray magnetic field produced by both strained and pristine CrSBr ribbons under increasing in-plane applied magnetic field. We apply the field $\mu_0H_x$ nearly along the $b$-axis of each ribbon. As revealed from AFM measurements, shown in Figure~\ref{fig2}a and c, the pristine ribbon is 18-layers-thick with an additional narrow single layer along the right edge, like the strained ribbon. Before imaging, we initialize the magnetic state of both ribbons by saturating the magnetization along the $b$-axis with an applied field $\mu_0H_x=-500$~mT. We then sweep the field to zero before incrementally increasing it and imaging $B_z^\text{ac} (x,y)$ up to $\mu_0H_x=600$~mT. In low applied field $\mu_0 H_x < 100$~mT, both ribbons remain in an AF state, with their only magnetic contrast due to the uncompensated strip near each ribbon's right edge, as observed in Figures~\ref{fig2}b and d. Further increases of $H_x$, however, result in a different evolution of the magnetic state in each ribbon. 

In the strained ribbon, as shown in Figure~\ref{fig2}b, a magnetization reorientation process occurs at $\mu_0H_x=203$~mT producing a clear signature in $B_z^\text{ac}$, corresponding to uncompensated magnetization pointing along the applied field. Although most of the ribbon is covered by this contrast, a section near the top of the image and an area along the right edge of the ribbon remain in the AF configuration. The wavy vertical features delineate the boundary between uncompensated and AF regions and are discussed, along with other features, in the next section. Upon increasing $\mu_0H_x$, the uncompensated region gradually expands throughout the ribbon, covering it almost completely at $\mu_0H_x=225$~mT (see Supplementary Figure 1). Here, the magnitude of $B_z^\text{ac} (x,y)$ corresponds to that expected for two uncompensated magnetic layers (see Supplementary Figure 5). From this image, we infer that the ribbon's bottom layer flips along the applied field, resulting in two full uncompensated layers (bottom and top) aligned along the field, while most of the single narrow uncompensated strip along the right edge remains oriented against the applied field. We assume the bottom layer to flip, because surface layers couple to only one adjacent layer, resulting in a lower energetic cost for reorientation than interior layers~\cite{cenker_strain-programmable_2023, wilson_interlayer_2021, yao_multiple_2023}, while the top layer is already oriented along the field opposing the narrow uncompensated strip. By $\mu_0H_x=250$~mT, both the bottom layer and the narrow single-layer strip have fully oriented along the field. Upon further increase of the applied field, at $\mu_0 H_x = 305$~mT, a small region of uncompensated magnetization forms where the tensile strain is largest, as shown in Figure~\ref{fig2}b. The magnetic contrast corresponds to that of two further uncompensated layers, indicating the flipping of an interior layer. As shown in Figure~\ref{fig2}b, subsequent increases in $H_x$ -- especially between $\mu_0 H_x = 310$ and 350~mT -- result in the expansion of this contrast along the left edge of the ribbon, accompanied by discrete increases in its magnitude, which emerge from the region of largest tensile strain. The behavior is consistent with additional layers orienting along the field, starting from the region of largest tensile strain. Eventually, the polarized region propagates across the entire ribbon, saturating around $\mu_0 H_x=400$~mT. A complete series of $B_z^\text{ac} (x,y)$ maps illustrating the magnetic evolution during both the upward and downward sweeps of $H_x$ can be found in Supplementary Figures 1 and 2, respectively. 

In the pristine ribbon, as shown in Figure~\ref{fig2}d, no magnetization reorientation is visible up to $\mu_0 H_x=100$~mT, at which point the narrow uncompensated strip along the right edge flips along the field (see Supplementary Figure 3). At $\mu_0 H_x=264$~mT, an abrupt transition occurs throughout the sample. As in the strained case, the contrast corresponds to two uncompensated layers, indicating the flipping of the bottom layer. As shown in Figure~\ref{fig2}d, this configuration persists until $\mu_0 H_x=330$~mT, at which point contrast corresponding to the reorientation of an additional flipped layer appears near the top end of the ribbon. At $\mu_0 H_x=350$~mT, similar contrast appears near the bottom end, followed by its expansion from both ends towards the middle of the ribbon. In this process, the magnitude of the contrast increases, indicating the reorientation of additional layers. The ribbon eventually saturates around $ \mu_0 H_x=400$~mT. Note that the strong magnetic contrast visible in the top right corner of the images in Figure~2d originates from a nearby thick flake. A full set of  $B_z^\text{ac} (x,y)$ maps for the pristine ribbon can be found in Supplementary Figures~3 and 4. 
%In comparison to the strained ribbon, the reorientation observed in the unstrained ribbon occurs in a narrower field range.

\section{Discussion}\label{sec6}

\subsection{Influence of strain on magnetic reversal}\label{sec7}

\begin{figure}[]%
\centering
\includegraphics[width=1\textwidth]{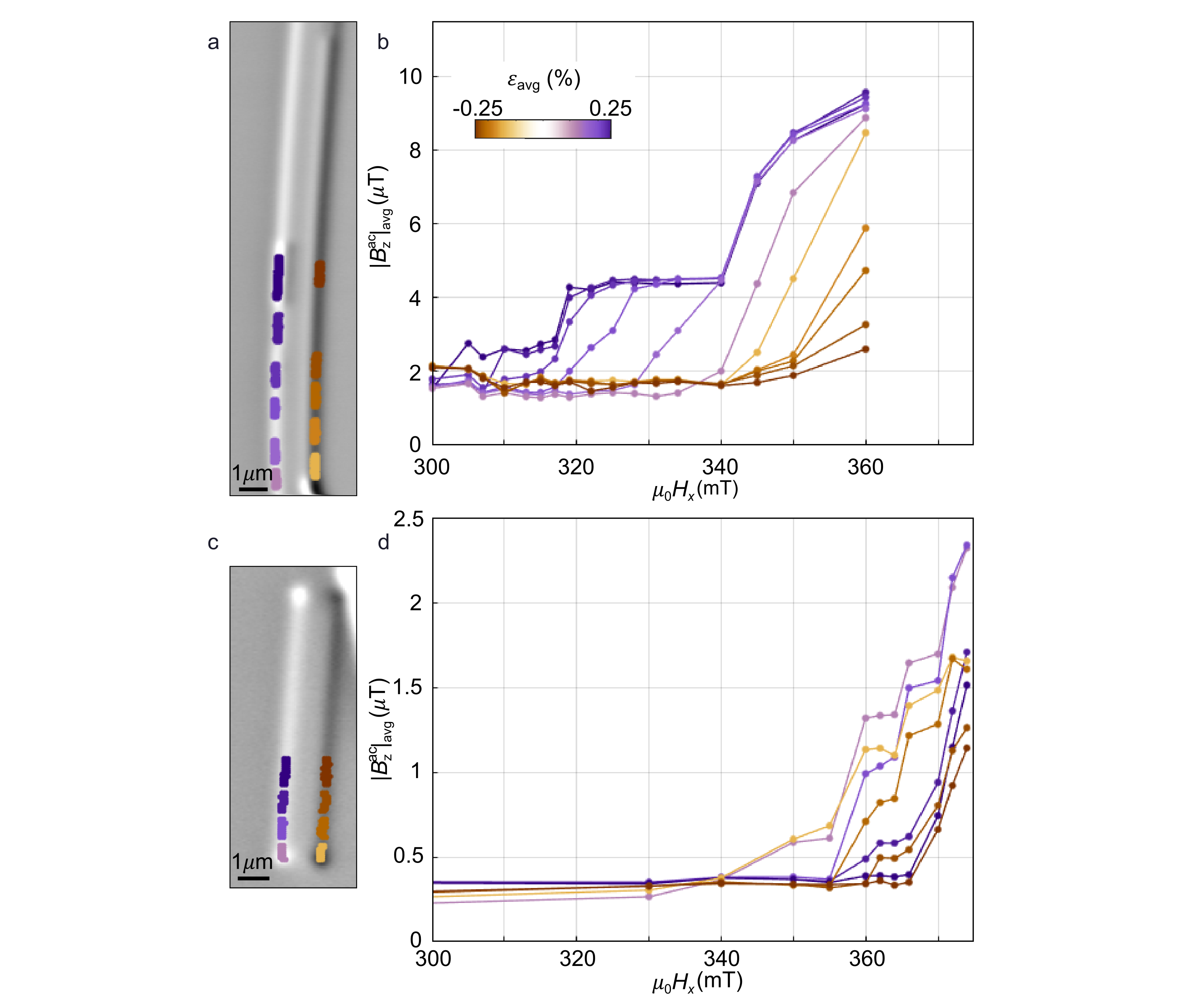}
\caption{\textbf{Strain-dependent reversal.} (a) $B_z^\text{ac}(x, y)$ map of the strained ribbon with $\mu_0H_x=305$~mT. (b) $|B_z^\text{ac}|_\text{avg}$ plotted as a function of increasing $\mu_0H_x$ averaged over the color-coded areas in (a) corresponding to different average $a$-axis strain shown in \% in the legend. (c) $B_z^\text{ac}(x, y)$ map of the pristine ribbon with $\mu_0H_x=350$~mT and (d) $|B_z^\text{ac}|_\text{avg}$ plotted as a function of increasing $\mu_0H_x$ averaged over the color-coded areas presented (c). }\label{fig3}
\end{figure}

The inhomogeneously strained ribbon, initially in its AF state, is polarized layer-by-layer by a magnetic field applied along its easy-axis. As shown in Figure~\ref{fig2}b, this reorientation begins in regions of the largest tensile strain, which weakens the interlayer AF coupling, eventually expanding into regions of compressive strain, which enhances the AF interaction. Tuning the interlayer coupling more strongly influences the magnetic reversal of interior layers than surface layers, due to their coupling with an additional adjacent layer. This is evidenced by the more abrupt polarization of the surface layer in Figure~\ref{fig2}b. In the pristine ribbon, the same reversal process, shown in Figure~\ref{fig2}d, begins at a higher field, i.e.\ $\mu_0 H_x = 264$ instead of $203$~mT, because of the lack of tensile strain weakening the AF interaction. The reversal also occurs over a narrower field range, due to less strain inhomogeneity compared to the strained ribbon. Only the narrow single-layer found along the right edge of both ribbons is observed to flip in the pristine ribbon first: at $\mu_0 H_x=100$~mT in the pristine ribbon compared to $225$~mT in the strained ribbon. In the strained ribbon, this strip experiences compressive strain, enhancing the interlayer AF interaction and thus increasing its switching field relative to the strip on the pristine ribbon.

The effect of strain on magnetic reversal is made apparent by plotting the local magnetic response of different regions of the sample. 
In Figure~\ref{fig3}a, we highlight regions in the strained ribbon with different average strain along the $a$-axis and, in Figure~\ref{fig3}b, we plot their average $|B_z^\text{ac}|$ as a function of $\mu_0 H_x$.
A strong correlation is evident between the magnetic switching behavior of different regions of the ribbon -- as indicated by each local $|B_z^\textbf{ac}|_\text{avg} (H_x)$ curve -- and their corresponding strain. Specifically, the field required to initiate the reorientation process increases with increasing compressive strain: the regions with the most tensile strain switch first, while those with the most compressive switch last. The reorientation of the first two interior layers is clearly visible in Figure~\ref{fig3}b as two discrete steps in the stray field from the regions corresponding to the most tensile strain. Subsequent steps or similar behavior in less strained regions occur in a narrower range of $H_x$ and are therefore not resolved over the measured applied field interval. 

Figures~\ref{fig3}c and d highlight different regions of the unstrained ribbon and show their local magnetic response. The $|B_z^\textbf{ac}|_\text{avg} (H_x)$ curves show significantly less variation in switching fields than in the strained ribbon. Although steps in $|B_z^\text{ac}|_\text{avg}(H_x)$, which correspond to the reorientation of individual layers, are recognizable in some regions, most occur over such small ranges in $H_x$ that the measurement does not resolve them. Despite the absence of intentional strain, the $|B_z^\text{ac}|_\text{avg}(H_x)$ curves show a noticeable trend of increasing switching field from the bottom and top edges towards the center of the pristine ribbon. Given the sensitivity of magnetic reversal to small strain gradients in the strained ribbon (down to fractions of a percent per micrometer), this spatial dependence could be the result of unintentional strain inhomogeneity.  Although the switching behavior near the top end of the ribbon could be influenced by the stray field of a neighboring flake, seen in the top right of Figure~\ref{fig3}c, similar behavior observed near the bottom end of the ribbon makes this possibility unlikely. Given that magnetic reorientation starts at the ribbon's ends and expands towards the middle along the long axis, the strongest strain gradients likely occur near the ends. Crystal growth, exfoliation, or the process of transferring the ribbon to the substrate may all induce inhomogeneous strain in the sample~\cite{klein_control_2022, rizzo_visualizing_2022, cenker_strain-programmable_2023}. %Nevertheless, we cannot exclude that other mechanisms do not also play a role in producing the spatially inhomogeneous reversal, including the presence of defects or impurities. 
%Both flakes show hysteretic behavior when comparing sweeps of $ \mu_0 H_x$ up and down, as discussed in Supplementary Note~2 and presented in Supplementary Figures~5.
% main arguments:
% - very clear correlation between strain and switching behavior (fig3c)
 %- the difference of the reversal procedure for the tensile and compressive sides of the flake is even more apparent in fig3c where Bzac is averaged along the two edges  (dB roughly 30mT). 

% For comparison, we perform the same analysis on the pristine flake.

%- transition starts later (fields comparable to the compressive region of the strained flake)
%- faster switching 
%- smaller variations among different areas
%- starts from the edges and expands to the middle
%- similar behavior between the two edges (already indicated by fig 3e but even more apparent in 3f
%-note:  the behavior of the top side of the pristine flake could be influenced by the stray field of the neighboring flake but the bottom edge presents similar behavior.

\subsection{Comparison to simulations}\label{sec8}

\begin{figure}[]%
\centering
\includegraphics[width=1\textwidth]{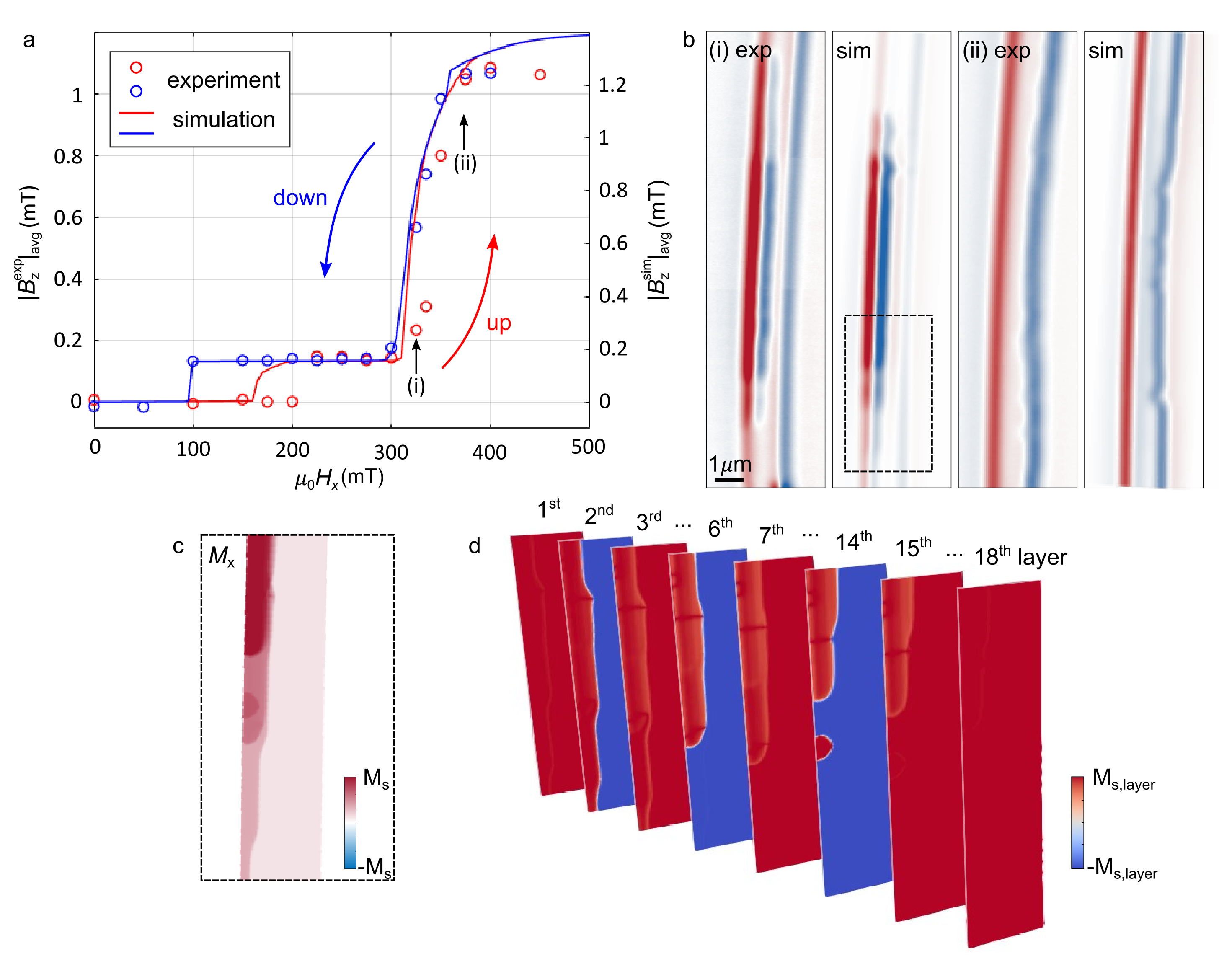}
\caption{\textbf{Visualizing the magnetization configuration during reversal.} (a) A comparison between experimental and simulated magnetic hysteresis for the strained ribbon. The  vertical axis corresponds to $|B_z|$ averaged over the entire imaging area shown in each panel of (b), which includes the central part of the ribbon. (b) The domains observed in the experimental $B_z^\text{ac} (x,y)$ maps are compared to the ones produced by the simulated $\frac{\text{d}B_z(x, y)}{\text{d}z}$ under (i) $\mu_0H_x=322$~mT in the experiment and $\mu_0H_x=315$~mT in the simulation and (ii) $\mu_0H_x=375$~mT for both experiment and simulation. (c) The total magnetization along the $x$-axis, $M_x (x,y)$, for the region indicated by the black dotted box in (b). $M_\text{s}$ represents the total saturation magnetization across all 18 layers.  The magnetization of this area is also plotted for selected layers in (d). $M_\text{s, layer}$ denotes the saturation magnetization of a single layer.}\label{fig4}
\end{figure}

To visualize the magnetic configurations underlying our measured stray-field maps and to understand the character of the magnetic reversal, we perform micromagnetic simulations based on the Landau-Liftshitz-Gilbert formalism~\cite{vansteenkiste_design_2014, exl_labontes_2014}. We simulate the main part of the 18-layer-thick bent ribbon using its geometry as determined from AFM images and known material parameters. We also implement a locally varying inter-layer exchange stiffness $A_\text{ex, inter}$, based on the map of applied $a$-axis strain, shown in Figure~\ref{fig1}d, and the theoretical dependence of $A_\text{ex, inter}$ on strain~\cite{bo_calculated_2023}, as discussed in the Methods.
%The resulting $A_{\text{ex, inter}} (x,y)$ map and further details that were used for this calculation can be found in Supplementary Figure 4. 
Figures~\ref{fig4}a and b show how simulations of the strained ribbon reproduce both the observed magnetic hysteresis and the measured $B_z^\text{ac} (x,y)$ maps as the strained ribbon is polarized by a field applied along its easy axis. Simulations were also carried out assuming both position-dependent magnetocrystalline anisotropy of the $a$-axis ($K_a$) and strain-dependent intralayer exchange stiffness ($A_\text{ex, intra}$), but neither reproduced the measurements. 

%In the measurements, the remaining layers flip at different applied fields in different regions experiencing different strains, as indicated by the $B_z^\text{ac}(x,y)$ contrast increasing in steps until saturation. This process may be the result of strain gradients along the thickness of the ribbon~\cite{cenker_strain-programmable_2023}. Although simulations do not include such gradients, on some iterations they reproduce this step-wise reorientation and on others they show the inner layers flipping in a single reorientation event. This variability in outcomes indicates the closeness in energy of these transitions and their potential for being influenced by small inhomogeneities and defects in the material.

The simulations confirm that the first layer to flip along the applied field, seen in both measured and simulated hysteresis curves as a step around $\mu_0 H = 200$~mT in Figure~\ref{fig4}a, is indeed a surface layer (bottom or top, depending on the initial AF configuration with respect to the applied field). They also explain specific features in maps of $B_z^\text{ac} (x,y)$ observed during reversal. In particular, they reproduce both the initiation of magnetic reorientation in the region of largest tensile strain and its expansion out from this area. As seen in Figure~\ref{fig4}b and Supplementary Figure 6, even the wavy and discontinuous features observed in $B_z^\text{ac} (x,y)$ maps at the borders of polarized domains appear in simulated images of similar configurations. These features correspond to boundaries extending through the thickness of the ribbon between regions with different numbers of uncompensated layers, as seen in Figure~\ref{fig4}c and d. These boundaries and their positions arise because of the spatially varying strain induced by the bending process and its effect on $A_\text{ex, inter}$. Since the simulations only consider this spatial dependence in the  sample, we conclude that these boundaries are not due to other inhomogeneities, imperfections, or defects in the ribbon itself.  The simulations also reproduce layer-by-layer switching without accounting for any vertical strain gradients as suggested in~\cite{cenker_strain-programmable_2023}. To minimize the demagnetization field during polarization and because interlayer coupling is much weaker than intralayer, the flake flips one layer at a time rather than all at once. Moreover, according to the simulations, the domain walls between the oppositely polarized domains are Néel-type, similar to what was observed by Zur et al \cite{zur_magnetic_2023}.

Similar simulations of the pristine ribbon using its measured geometry, thickness, and uniform parameters -- including $A_\text{ex, inter}$ -- do not reproduce the observed reorientation behavior. In particular, simulations do not show reversal proceeding through spatially confined domains of uncompensated magnetization. Rather, the entire ribbon goes through a  layer-by-layer reorientation over a much narrower range of $H_x$ than in the measurements. Reorientation that begins at the top and bottom ends of the ribbon can therefore not be explained simply by the geometry of the sample and the resulting inhomogeneity in its demagnetization field. However, the simulations do display behavior similar to the measurements upon the introduction of a spatially varying $A_\text{ex, inter}$ near the top and bottom of the ribbon, supporting the hypothesis of unintentional strain at the ends. Nevertheless, we cannot exclude that other inhomogeneities besides strain contribute to this behavior.

\section{Conclusions}\label{sec9}
Using a nanometer-scale SQUID-on-lever probe, we  map the local magnetic stray field produced by intentionally strained and pristine ribbons of CrSBr as a function of field applied along its easy axis. These images show how both strain and strain inhomogeneity affect the way that the system is driven from a remanent AF state to a F state by an increasing applied field. In particular, a micromagnetic model of the system, which takes into account the spatially varying strain and its coupling to the interlayer exchange stiffness, matches the experimentally observed behavior, including the formation and evolution of domains. 

We show both the unambiguous coupling of strain to magnetic reversal and the ability to locally tune the interlayer exchange stiffness via the application of strain. Such local control over the magnetic properties of a 2D magnet could be used in devices to program the formation of domains or to spatially confine magnetic reversal. Measurements also suggest that small strain gradients -- in some cases unintentional -- can affect the sample's magnetic hysteresis. These gradients could be produced by the exfoliation and deposition process, and can result in an inhomogeneous evolution of magnetic hysteresis throughout the sample. Sensitivity to unintentional strain must therefore be considered in the design of any future magnetic devices involving CrSBr or similar materials. The effects of unintentional strain may explain inhomogeneities observed in previous spatially resolved~\cite{tschudin_imaging_2024, yu_direct_2024, tabataba-vakili_doping-control_2024} and transport~\cite{boix-constant_multistep_2024} measurements.

\section{Methods}\label{sec11} 
\subsection{CrSBr ribbons}

CrSBr single crystals are grown using the chemical vapor transport technique as described in Scheie et al.~\cite{scheie_spin_2022}. Nominally unstrained CrSBr ribbons are obtained by mechanical exfoliation of bulk crystals onto SiO$_2$/Si chips. A polycarbonate/polydimethylsiloxane (PC-PDMS) transfer stamp~\cite{wang_one-dimensional_2013} is used to pick up exfoliated hBN, CrSBr, and a metallized graphite micro-manipulator in sequence. See Kapfer et al.~\cite{kapfer_programming_2023} for micro-manipulator preparation. Subsequently, the completed stack is flipped onto a target SiO$_2$/Si chip such that the CrSBr flake and micro-manipulator are exposed. The micro-manipulator is maneuvered to deform the CrSBr flake using an AFM operated in contact mode and the ``Nanoman'' software (Bruker). 

We measure the thickness of different regions of the sample using AFM. We convert this measurement to a number of atomic layers using a thickness of 0.78~$\text{nm}/\text{layer}$ for all layers~\cite{lee_magnetic_2021, goser_magnetic_1990} except the first layer, which is in direct contact with the hexagonal boron nitride and is assumed to be 1.1-nm-thick based on the measurement.

\subsection{SQUID-on-lever}

We fabricate the SOL scanning probes by sputtering Nb on a AFM cantilever and patterning its apex via focused-ion-beam milling, as described by Wyss et al.~\cite{wyss_magnetic_2022}. Each hybrid AFM/SQUID sensor is characterized and operated at 4.2~K in a semi-voltage biased circuit, in which the current response $I_{\text{SQUID}}$ is measured by a series SQUID array amplifier (Magnicon). Two different SOL sensors were used for this study. Their effective diameters of 175~nm and  185~nm are extracted from their corresponding quantum interference patterns. The sensors have an AC magnetic field sensitivity exceeding $S^{1/2}_{B} = 1$~$\mu T 
 ~\text{Hz}^{-1/2}$ in the white-noise regime. 

\subsection{Magnetic imaging}

Magnetic imaging is carried out in a custom-built scanning probe microscope operating under high vacuum within a \(^4\text{He}\) cryostat. The SOL scanning probe is capable of simultaneously performing non-contact AFM and scanning SQUID microscopy. The cantilever is excited to an amplitude of $\Delta z = 16$~nm at its fundamental mechanical resonance frequency by a piezo-electric actuator driven by a phase-locked loop. Displacement oscillations are detected using a fiber-optic interferometer. Since $I_{\text{SQUID}}$ is proportional to magnetic flux, this response provides a measure of the local magnetic field threading through the SQUID loop. Note that the axis of the SQUID is tilted by $10^\circ$ with respect to the $z$-axis, because the SOL is scanned in a standard AFM configuration designed to easily approach the sensor tip without touching the rest of the cantilever body. As a result, both $B_z (x, y)$ and $B_z^\text{ac} (x,y)$ are measured with a $10^\circ$ tilt with respect to the $z$-axis. Simulated maps of the same quantities take this tilt into account.

We measure the response of $I_\text{SQUID}$ as a function of the out-of-plane applied magnetic field $\mu_0H_z$ before and after each scan for a field range larger than that produced by the ribbon. By scanning the sample using a scanning probe controller (Specs) and piezoelectric actuators (Attocube) at a constant tip-sample spacing of 250~nm, we map $B_z (x, y)$ and $B_z^\text{ac} (x,y)$ in a plane above the sample. $B_z^\text{ac} \propto d B_z / d z$ is measured by demodulating the SQUID-on-lever response at the cantilever oscillation frequency. Due to spectral filtering, the resulting signal contains less noise than DC measurements of $B_z$. Spatial resolution is limited by the tip-sample spacing and by the $\sim$180~nm effective diameter of the SQUID loop.

In all imaging experiments, we apply a small out-of-plane field between $\mu_0 |H_z|=10$ and 20~mT. This out-of-plane field ensures that the SQUID sensor, whose magnetic sensitivity is field-dependent, is sensitive enough to measure the sample's stray fields. Each layer's strong in-plane anisotropy should ensure that the magnetization remains nearly unaffected by the small $H_z$~\cite{telford_layered_2020}.

\subsection{Micromagnetic simulations}

We simulate the magnetization configuration of the CrSBr ribbons using the \textit{Mumax$^3$} software package~\cite{vansteenkiste_design_2014, exl_labontes_2014, leliaert_adaptively_2017}, which uses the Landau-Lifshitz-Gilbert micromagnetic formalism with finite-difference discretization. To mimic the layered structure of CrSBr, we set the thickness of a cell in the finite-difference mesh to the thickness of a CrSBr layer. We use literature values to set the saturation magnetization $M_\text{sat} =36\, \mu_\text{B}/\text{nm}^2$~\cite{tschudin_imaging_2024, lopez-paz_dynamic_2022}, the magnetocrystalline anisotropy along the $a$- and $b$-axis $K_\text{u,a}$, $K_\text{u,b}$~\cite{tschudin_imaging_2024}, and the intralayer exchange interactions J$_1$, J$_2$, J$_3$~\cite{scheie_spin_2022}. These exchange interactions are used to estimate the intralayer exchange stiffness $A_\text{ex, intra}$~\cite{skomski_permanent_1999}.  We estimate the interlayer exchange stiffness $A_\text{ex, inter}$ and its dependence on strain from theoretical calculations~\cite{bo_calculated_2023}.  For the values of strain considered in this study, $A_\text{ex, inter}$ depends linearly on strain. As a starting point, we simulate the magnetic hysteresis of a cube of $200 \text{ nm}\times 200 \text{ nm}\times 200 \text{ nm}$ using a cell size of $4 \text{ nm}\times 4 \text{ nm}\times 0.8 \text{ nm}$ and periodic boundary conditions. We compare the results to bulk hysteresis measurements~\cite{goser_magnetic_1990, lopez-paz_dynamic_2022, pei_surfacesensitive_2024} along the $a$- $b$- and $c$-axis and modify the values to 
 $K_\text{u, a} = 72$~kJ/m$^3$, $K_\text{u, b} = 127$~kJ/m$^3$, and $A_\text{ex, inter} = -23$~fJ/m in order to optimize the match. Changes in $A_\text{ex, intra}$ do not have a significant influence on the simulated hysteresis (within $1-10$~pJ/m range). Therefore, we choose $A_\text{ex, intra}=8$~pJ/m to increase the exchange length  $l_\text{ex}$, allowing us to simulate a larger area. 

For the simulation of the strained ribbon, the geometry and thickness in number of layers is determined from AFM measurements. The simulations assume a uniform thickness of 18 layers, ignoring the narrow region along the right edge of the ribbon consisting of a single-layer strip followed by a step with decreased thickness (14 layers). The structure is discretized into cells of size $8 \text{ nm}\times 8 \text{ nm}\times 0.8 \text{ nm}$. Starting from the value of  $A_\text{ex, inter}$ determined for the simulation of bulk material and its theoretical strain dependence, we adjust this dependence to optimally reproduce both the measured magnetic hysteresis and the measured stray field patterns, resulting in $A_\text{ex, inter}=(16.3 \varepsilon (\%) -23)$~fJ/m, where $\varepsilon$ is the strain along the $a$-axis.  This linear dependence agrees roughly with previous observations, predicting an AF to F transition at 1.5\% strain compared to the observed 1.3\%~\cite{cenker_reversible_2022}. The spatial variation of strain is implemented by dividing the ribbon into 110 regions of uniform $A_\text{ex, inter}$, corresponding to the average value in that region. 

%Simulations with uniform  $A_\text{ex, inter}$ within the flake do not reproduce the magnetic domains observed experimentally. Introducing strain dependent $K_\text{u, a}$ and/or $A_\text{ex, intra}$ while keeping $A_\text{ex, inter}$ constant affected the magnetic hysteresis of the simulated flake, but still did not result in the formation of the observed domains. These findings suggest that non-uniformities in $A_\text{ex, inter}$ produced by strain variations are crucial for the formation and evolution of the observed domains under applied $\mu_0H_x$ while possible variations in magnetocrystalline anisotropies or intralayer exchange interactions play a minor role. 
%Some material parameters had to be adjusted to best reproduce the behavior observed in our flake.
Simulated $B_z(x, y)$ and $B_z^\text{ac}(x,y)$ are calculated from the magnetization maps generated by \textit{Mumax$^3$} at a height of 250~nm above the sample, corresponding to our SQUID-sample distance. $B_z^\text{ac}(x,y)$ is calculated assuming $\text{d}z=\SI{16}{\nano\meter}$, which corresponds to the oscillation amplitude of the cantilever. Finally, we apply a Gaussian blurring of $2\sigma=\SI{192}{\nano\meter}$ to approximate the point-spread function of the SQUID sensor.

\pagebreak

% =========================================================================
\bmhead{Data Availability}

The data supporting the findings of this study are available from the authors upon reasonable request.

% The data supporting the findings of this study are available on the Zenodo repository at https://doi.org/10.5281/zenodo.10718021.

\bmhead{Author Contributions}

M.A.T., K.B., and A.V. conceived the project. D.G.C. synthesized the CrSBr single-crystals. A.D. fabricated the sample. D.J. fabricated the SOL sensors with support from K.B and A.V. K.B. and A.V. performed the experiment with support from D.J. A.V. and K.B. analyzed the data. A.V. performed the micromagnetic simulations with support from B.G. M.P., A.V., and K. B. wrote the manuscript. M.P. provided supervision and led the collaboration. All authors discussed the results and commented on the manuscript.

\bmhead{Competing interests}
The authors declare no competing interests.

\bmhead{Acknowledgments}

We thank Sascha Martin and his team in the machine shop of the Department of Physics at the University of Basel for their role in building the scanning probe microscope. We also thank Monica Schönenberger from the Nano Imaging Lab of the Swiss Nanoscience Institute for help with AFM. Calculations were performed at sciCORE (http://scicore.unibas.ch/) scientific computing center at the University of Basel. We acknowledge support of the European Commission under H2020 FET Open grant "FIBsuperProbes" (Grant No. 892427), the SNF under Grant No. 200020-207933, and the Canton Aargau. Synthetic work at Columbia was supported by the National Science Foundation (NSF) through the Columbia Materials Science and Engineering Research Center on Precision-Assembled Quantum Materials (DMR-2011738). A.D.\ acknowledges support from the Simons Foundation Society of Fellows program (Grant No. 855186).

\pagebreak

%\bibliography{CrSBr}

\begin{thebibliography}{10}
\expandafter\ifx\csname url\endcsname\relax
  \def\url#1{\burl{#1}}\fi
\expandafter\ifx\csname urlprefix\endcsname\relax\def\urlprefix{URL }\fi
\providecommand{\bibinfo}[2]{#2}
\providecommand{\eprint}[2][]{\url{#2}}
\providecommand{\doi}[1]{\url{https://doi.org/#1}}
\bibcommenthead

\bibitem{li_tuning_2021}
\bibinfo{author}{Li, D.}, \bibinfo{author}{Li, S.}, \bibinfo{author}{Zhong, C.}
  \& \bibinfo{author}{He, J.}
\newblock \bibinfo{title}{Tuning magnetism at the two-dimensional limit: a
  theoretical perspective}.
\newblock \emph{\bibinfo{journal}{Nanoscale}} \textbf{\bibinfo{volume}{13}},
  \bibinfo{pages}{19812--19827} (\bibinfo{year}{2021}).

\bibitem{verzhbitskiy_controlling_2020}
\bibinfo{author}{Verzhbitskiy, I.~A.} \emph{et~al.}
\newblock \bibinfo{title}{Controlling the magnetic anisotropy in
  {Cr}$_{\textrm{2}}${Ge}$_{\textrm{2}}${Te}$_{\textrm{6}}$ by electrostatic
  gating}.
\newblock \emph{\bibinfo{journal}{Nature Electronics}}
  \textbf{\bibinfo{volume}{3}}, \bibinfo{pages}{460--465}
  (\bibinfo{year}{2020}).

\bibitem{zhuo_manipulating_2021}
\bibinfo{author}{Zhuo, W.} \emph{et~al.}
\newblock \bibinfo{title}{Manipulating {Ferromagnetism} in {Few}-{Layered}
  {Cr}$_{\textrm{2}}${Ge}$_{\textrm{2}}${Te}$_{\textrm{6}}$}.
\newblock \emph{\bibinfo{journal}{Advanced Materials}}
  \textbf{\bibinfo{volume}{33}}, \bibinfo{pages}{2008586}
  (\bibinfo{year}{2021}).

\bibitem{siskins_magnetic_2020}
\bibinfo{author}{Šiškins, M.} \emph{et~al.}
\newblock \bibinfo{title}{Magnetic and electronic phase transitions probed by
  nanomechanical resonators}.
\newblock \emph{\bibinfo{journal}{Nature Communications}}
  \textbf{\bibinfo{volume}{11}}, \bibinfo{pages}{2698} (\bibinfo{year}{2020}).

\bibitem{wang_strainsensitive_2020}
\bibinfo{author}{Wang, Y.} \emph{et~al.}
\newblock \bibinfo{title}{Strain‐{Sensitive} {Magnetization} {Reversal} of a
  van der {Waals} {Magnet}}.
\newblock \emph{\bibinfo{journal}{Advanced Materials}}
  \textbf{\bibinfo{volume}{32}}, \bibinfo{pages}{2004533}
  (\bibinfo{year}{2020}).

\bibitem{goser_magnetic_1990}
\bibinfo{author}{Göser, O.}, \bibinfo{author}{Paul, W.} \&
  \bibinfo{author}{Kahle, H.}
\newblock \bibinfo{title}{Magnetic properties of {CrSBr}}.
\newblock \emph{\bibinfo{journal}{Journal of Magnetism and Magnetic Materials}}
  \textbf{\bibinfo{volume}{92}}, \bibinfo{pages}{129--136}
  (\bibinfo{year}{1990}).

\bibitem{telford_layered_2020}
\bibinfo{author}{Telford, E.~J.} \emph{et~al.}
\newblock \bibinfo{title}{Layered {Antiferromagnetism} {Induces} {Large}
  {Negative} {Magnetoresistance} in the van der {Waals} {Semiconductor}
  {CrSBr}}.
\newblock \emph{\bibinfo{journal}{Advanced Materials}}
  \textbf{\bibinfo{volume}{32}}, \bibinfo{pages}{2003240}
  (\bibinfo{year}{2020}).

\bibitem{telford_coupling_2022}
\bibinfo{author}{Telford, E.~J.} \emph{et~al.}
\newblock \bibinfo{title}{Coupling between magnetic order and charge transport
  in a two-dimensional magnetic semiconductor}.
\newblock \emph{\bibinfo{journal}{Nature Materials}}
  \textbf{\bibinfo{volume}{21}}, \bibinfo{pages}{754--760}
  (\bibinfo{year}{2022}).

\bibitem{cenker_reversible_2022}
\bibinfo{author}{Cenker, J.} \emph{et~al.}
\newblock \bibinfo{title}{Reversible strain-induced magnetic phase transition
  in a van der {Waals} magnet}.
\newblock \emph{\bibinfo{journal}{Nature Nanotechnology}}
  \textbf{\bibinfo{volume}{17}}, \bibinfo{pages}{256--261}
  (\bibinfo{year}{2022}).

\bibitem{cenker_strain-programmable_2023}
\bibinfo{author}{Cenker, J.} \emph{et~al.}
\newblock \bibinfo{title}{Strain-programmable van der {Waals} magnetic tunnel
  junctions}.
\newblock \emph{\bibinfo{journal}{arXiv:2301.03759}}  (\bibinfo{year}{2023}).

\bibitem{telford_designing_2023}
\bibinfo{author}{Telford, E.~J.} \emph{et~al.}
\newblock \bibinfo{title}{Designing {Magnetic} {Properties} in {CrSBr} through
  {Hydrostatic} {Pressure} and {Ligand} {Substitution}}.
\newblock \emph{\bibinfo{journal}{Advanced Physics Research}}
  \textbf{\bibinfo{volume}{2}}, \bibinfo{pages}{2300036}
  (\bibinfo{year}{2023}).

\bibitem{tabataba-vakili_doping-control_2024}
\bibinfo{author}{Tabataba-Vakili, F.} \emph{et~al.}
\newblock \bibinfo{title}{Doping-control of excitons and magnetism in few-layer
  {CrSBr}}.
\newblock \emph{\bibinfo{journal}{Nature Communications}}
  \textbf{\bibinfo{volume}{15}}, \bibinfo{pages}{4735} (\bibinfo{year}{2024}).

\bibitem{shcherbakov_raman_2018}
\bibinfo{author}{Shcherbakov, D.} \emph{et~al.}
\newblock \bibinfo{title}{Raman {Spectroscopy}, {Photocatalytic} {Degradation},
  and {Stabilization} of {Atomically} {Thin} {Chromium} {Tri}-iodide}.
\newblock \emph{\bibinfo{journal}{Nano Letters}} \textbf{\bibinfo{volume}{18}},
  \bibinfo{pages}{4214--4219} (\bibinfo{year}{2018}).

\bibitem{wu_degradation_2022}
\bibinfo{author}{Wu, Y.} \emph{et~al.}
\newblock \bibinfo{title}{Degradation {Effect} and {Magnetoelectric}
  {Transport} {Properties} in {CrBr}$_{\textrm{3}}$ {Devices}}.
\newblock \emph{\bibinfo{journal}{Materials}} \textbf{\bibinfo{volume}{15}},
  \bibinfo{pages}{3007} (\bibinfo{year}{2022}).

\bibitem{mastrippolito_emerging_2021}
\bibinfo{author}{Mastrippolito, D.} \emph{et~al.}
\newblock \bibinfo{title}{Emerging oxidized and defective phases in
  low-dimensional {CrCl}$_{\textrm{3}}$}.
\newblock \emph{\bibinfo{journal}{Nanoscale Advances}}
  \textbf{\bibinfo{volume}{3}}, \bibinfo{pages}{4756--4766}
  (\bibinfo{year}{2021}).

\bibitem{lee_magnetic_2021}
\bibinfo{author}{Lee, K.} \emph{et~al.}
\newblock \bibinfo{title}{Magnetic {Order} and {Symmetry} in the {2D}
  {Semiconductor} {CrSBr}}.
\newblock \emph{\bibinfo{journal}{Nano Letters}} \textbf{\bibinfo{volume}{21}},
  \bibinfo{pages}{3511--3517} (\bibinfo{year}{2021}).

\bibitem{tschudin_imaging_2024}
\bibinfo{author}{Tschudin, M.~A.} \emph{et~al.}
\newblock \bibinfo{title}{Imaging nanomagnetism and magnetic phase transitions
  in atomically thin {CrSBr}}.
\newblock \emph{\bibinfo{journal}{Nature Communications}}
  \textbf{\bibinfo{volume}{15}}, \bibinfo{pages}{6005} (\bibinfo{year}{2024}).

\bibitem{bo_calculated_2023}
\bibinfo{author}{Bo, X.}, \bibinfo{author}{Li, F.}, \bibinfo{author}{Xu, X.},
  \bibinfo{author}{Wan, X.} \& \bibinfo{author}{Pu, Y.}
\newblock \bibinfo{title}{Calculated magnetic exchange interactions in the van
  der {Waals} layered magnet {CrSBr}}.
\newblock \emph{\bibinfo{journal}{New Journal of Physics}}
  \textbf{\bibinfo{volume}{25}}, \bibinfo{pages}{013026}
  (\bibinfo{year}{2023}).

\bibitem{diao_strain-regulated_2023}
\bibinfo{author}{Diao, Y.} \emph{et~al.}
\newblock \bibinfo{title}{Strain-regulated magnetic phase transition and
  perpendicular magnetic anisotropy in {CrSBr} monolayer}.
\newblock \emph{\bibinfo{journal}{Physica E: Low-dimensional Systems and
  Nanostructures}} \textbf{\bibinfo{volume}{147}}, \bibinfo{pages}{115590}
  (\bibinfo{year}{2023}).

\bibitem{yang_triaxial_2021}
\bibinfo{author}{Yang, K.}, \bibinfo{author}{Wang, G.}, \bibinfo{author}{Liu,
  L.}, \bibinfo{author}{Lu, D.} \& \bibinfo{author}{Wu, H.}
\newblock \bibinfo{title}{Triaxial magnetic anisotropy in the two-dimensional
  ferromagnetic semiconductor {CrSBr}}.
\newblock \emph{\bibinfo{journal}{Physical Review B}}
  \textbf{\bibinfo{volume}{104}}, \bibinfo{pages}{144416}
  (\bibinfo{year}{2021}).

\bibitem{wang_origin_2023}
\bibinfo{author}{Wang, B.} \emph{et~al.}
\newblock \bibinfo{title}{Origin and regulation of triaxial magnetic anisotropy
  in the ferromagnetic semiconductor {CrSBr} monolayer}.
\newblock \emph{\bibinfo{journal}{Nanoscale}} \textbf{\bibinfo{volume}{15}},
  \bibinfo{pages}{13402--13410} (\bibinfo{year}{2023}).

\bibitem{kapfer_programming_2023}
\bibinfo{author}{Kapfer, M.} \emph{et~al.}
\newblock \bibinfo{title}{Programming twist angle and strain profiles in {2D}
  materials}.
\newblock \emph{\bibinfo{journal}{Science}} \textbf{\bibinfo{volume}{381}},
  \bibinfo{pages}{677--681} (\bibinfo{year}{2023}).

\bibitem{wyss_magnetic_2022}
\bibinfo{author}{Wyss, M.} \emph{et~al.}
\newblock \bibinfo{title}{Magnetic, {Thermal}, and {Topographic} {Imaging} with
  a {Nanometer}-{Scale} {SQUID}-{On}-{Lever} {Scanning} {Probe}}.
\newblock \emph{\bibinfo{journal}{Physical Review Applied}}
  \textbf{\bibinfo{volume}{17}}, \bibinfo{pages}{034002}
  (\bibinfo{year}{2022}).

\bibitem{wilson_interlayer_2021}
\bibinfo{author}{Wilson, N.~P.} \emph{et~al.}
\newblock \bibinfo{title}{Interlayer electronic coupling on demand in a {2D}
  magnetic semiconductor}.
\newblock \emph{\bibinfo{journal}{Nature Materials}}
  \textbf{\bibinfo{volume}{20}}, \bibinfo{pages}{1657--1662}
  (\bibinfo{year}{2021}).

\bibitem{yao_multiple_2023}
\bibinfo{author}{Yao, F.} \emph{et~al.}
\newblock \bibinfo{title}{Multiple antiferromagnetic phases and magnetic
  anisotropy in exfoliated {CrBr}$_{\textrm{3}}$ multilayers}.
\newblock \emph{\bibinfo{journal}{Nature Communications}}
  \textbf{\bibinfo{volume}{14}}, \bibinfo{pages}{4969} (\bibinfo{year}{2023}).

\bibitem{klein_control_2022}
\bibinfo{author}{Klein, J.} \emph{et~al.}
\newblock \bibinfo{title}{Control of structure and spin texture in the van der
  {Waals} layered magnet {CrSBr}}.
\newblock \emph{\bibinfo{journal}{Nature Communications}}
  \textbf{\bibinfo{volume}{13}}, \bibinfo{pages}{5420} (\bibinfo{year}{2022}).

\bibitem{rizzo_visualizing_2022}
\bibinfo{author}{Rizzo, D.~J.} \emph{et~al.}
\newblock \bibinfo{title}{Visualizing {Atomically} {Layered} {Magnetism} in
  {CrSBr}}.
\newblock \emph{\bibinfo{journal}{Advanced Materials}}
  \textbf{\bibinfo{volume}{34}}, \bibinfo{pages}{2201000}
  (\bibinfo{year}{2022}).

\bibitem{vansteenkiste_design_2014}
\bibinfo{author}{Vansteenkiste, A.} \emph{et~al.}
\newblock \bibinfo{title}{The design and verification of {MuMax3}}.
\newblock \emph{\bibinfo{journal}{AIP Advances}} \textbf{\bibinfo{volume}{4}},
  \bibinfo{pages}{107133} (\bibinfo{year}{2014}).

\bibitem{exl_labontes_2014}
\bibinfo{author}{Exl, L.} \emph{et~al.}
\newblock \bibinfo{title}{{LaBonte}'s method revisited: {An} effective steepest
  descent method for micromagnetic energy minimization}.
\newblock \emph{\bibinfo{journal}{Journal of Applied Physics}}
  \textbf{\bibinfo{volume}{115}}, \bibinfo{pages}{17D118}
  (\bibinfo{year}{2014}).

\bibitem{zur_magnetic_2023}
\bibinfo{author}{Zur, Y.} \emph{et~al.}
\newblock \bibinfo{title}{Magnetic {Imaging} and {Domain} {Nucleation} in
  {CrSBr} {Down} to the {2D} {Limit}}.
\newblock \emph{\bibinfo{journal}{Advanced Materials}}
  \textbf{\bibinfo{volume}{35}}, \bibinfo{pages}{2307195}
  (\bibinfo{year}{2023}).

\bibitem{yu_direct_2024}
\bibinfo{author}{Yu, J.} \emph{et~al.}
\newblock \bibinfo{title}{Direct {Imaging} of {Antiferromagnet}‐{Ferromagnet}
  {Phase} {Transition} in van der {Waals} {Antiferromagnet} {CrSBr}}.
\newblock \emph{\bibinfo{journal}{Advanced Functional Materials}}
  \textbf{\bibinfo{volume}{34}}, \bibinfo{pages}{2307259}
  (\bibinfo{year}{2024}).

\bibitem{boix-constant_multistep_2024}
\bibinfo{author}{Boix-Constant, C.} \emph{et~al.}
\newblock \bibinfo{title}{Multistep magnetization switching in orthogonally
  twisted ferromagnetic monolayers}.
\newblock \emph{\bibinfo{journal}{Nature Materials}}
  \textbf{\bibinfo{volume}{23}}, \bibinfo{pages}{212--218}
  (\bibinfo{year}{2024}).

\bibitem{scheie_spin_2022}
\bibinfo{author}{Scheie, A.} \emph{et~al.}
\newblock \bibinfo{title}{Spin {Waves} and {Magnetic} {Exchange} {Hamiltonian}
  in {CrSBr}}.
\newblock \emph{\bibinfo{journal}{Advanced Science}}
  \textbf{\bibinfo{volume}{9}}, \bibinfo{pages}{2202467}
  (\bibinfo{year}{2022}).

\bibitem{wang_one-dimensional_2013}
\bibinfo{author}{Wang, L.} \emph{et~al.}
\newblock \bibinfo{title}{One-{Dimensional} {Electrical} {Contact} to a
  {Two}-{Dimensional} {Material}}.
\newblock \emph{\bibinfo{journal}{Science}} \textbf{\bibinfo{volume}{342}},
  \bibinfo{pages}{614--617} (\bibinfo{year}{2013}).

\bibitem{leliaert_adaptively_2017}
\bibinfo{author}{Leliaert, J.} \emph{et~al.}
\newblock \bibinfo{title}{Adaptively time stepping the stochastic
  {Landau}-{Lifshitz}-{Gilbert} equation at nonzero temperature:
  {Implementation} and validation in {MuMax3}}.
\newblock \emph{\bibinfo{journal}{AIP Advances}} \textbf{\bibinfo{volume}{7}},
  \bibinfo{pages}{125010} (\bibinfo{year}{2017}).

\bibitem{lopez-paz_dynamic_2022}
\bibinfo{author}{López-Paz, S.~A.} \emph{et~al.}
\newblock \bibinfo{title}{Dynamic magnetic crossover at the origin of the
  hidden-order in van der {Waals} antiferromagnet {CrSBr}}.
\newblock \emph{\bibinfo{journal}{Nature Communications}}
  \textbf{\bibinfo{volume}{13}}, \bibinfo{pages}{4745} (\bibinfo{year}{2022}).

\bibitem{skomski_permanent_1999}
\bibinfo{author}{Skomski, R.} \& \bibinfo{author}{Coey, J. M.~D.}
\newblock \emph{\bibinfo{title}{Permanent {Magnetism}}}
  (\bibinfo{publisher}{Institute of Physics Publishing}, \bibinfo{year}{1999}).

\bibitem{pei_surfacesensitive_2024}
\bibinfo{author}{Pei, F.} \emph{et~al.}
\newblock \bibinfo{title}{Surface‐{Sensitive} {Detection} of {Magnetic}
  {Phase} {Transition} in {Van} {Der} {Waals} {Magnet} {CrSBr}}.
\newblock \emph{\bibinfo{journal}{Advanced Functional Materials}}
  \textbf{\bibinfo{volume}{34}}, \bibinfo{pages}{2309335}
  (\bibinfo{year}{2024}).

\end{thebibliography}
%\addbibresource{CGT}

\end{document}

\title[ ]{Supplementary~Information: Imaging strain-controlled magnetic reversal in thin CrSBr }
%\title[Bagani et al.]{Controlling magnetic reversal via strain in CrSBr}

\author[1]{\fnm{Kousik} \sur{Bagani}}
\equalcont{These authors contributed equally to this work.}

\author[1]{\fnm{Andriani} \sur{Vervelaki}}
\equalcont{These authors contributed equally to this work.}

\author[1]{\fnm{Daniel} \sur{Jetter}}

\author[2,3]{\fnm{Aravind} \sur{Devarakonda}}

\author[1]{\fnm{Märta A.} \sur{Tschudin}}
\author[1]{\fnm{Boris} \sur{Gross}}
\author[4]{\fnm{Daniel G.} \sur{Chica}}
\author[1,5]{\fnm{David A.} \sur{Broadway}}
\author[3]{\fnm{Cory R.} \sur{Dean}}
\author[4]{\fnm{Xavier} \sur{Roy}}

\author[1]{\fnm{Patrick} \sur{Maletinsky}}

\author*[1,6]{\fnm{Martino} \sur{Poggio}}\email{martino.poggio@unibas.ch}

\affil*[1]{\orgdiv{Department of Physics}, \orgname{University of Basel}, \country{Switzerland}}

\affil[2]{\orgdiv{Department of Physics}, \orgname{Columbia University}, \country{USA}}

\affil[3]{\orgdiv{Department of Applied Physics and Applied Mathematics}, \orgname{Columbia University}, \country{USA}}

\affil[4]{\orgdiv{Department of Chemistry}, \orgname{Columbia University}, \country{USA}}

\affil[5]{\orgdiv{School of Science}, \orgname{RMIT University}, \country{Australia}}

\affil[6]{\orgdiv{Swiss Nanoscience Institute}, \orgname{University of Basel}, \country{Switzerland}}




\maketitle

\pagebreak

\section{Supplementary Note 1:\\ Magnetic Imaging}\label{secA1}

\begin{figure}[H]%
\centering
\includegraphics[width=\textwidth]{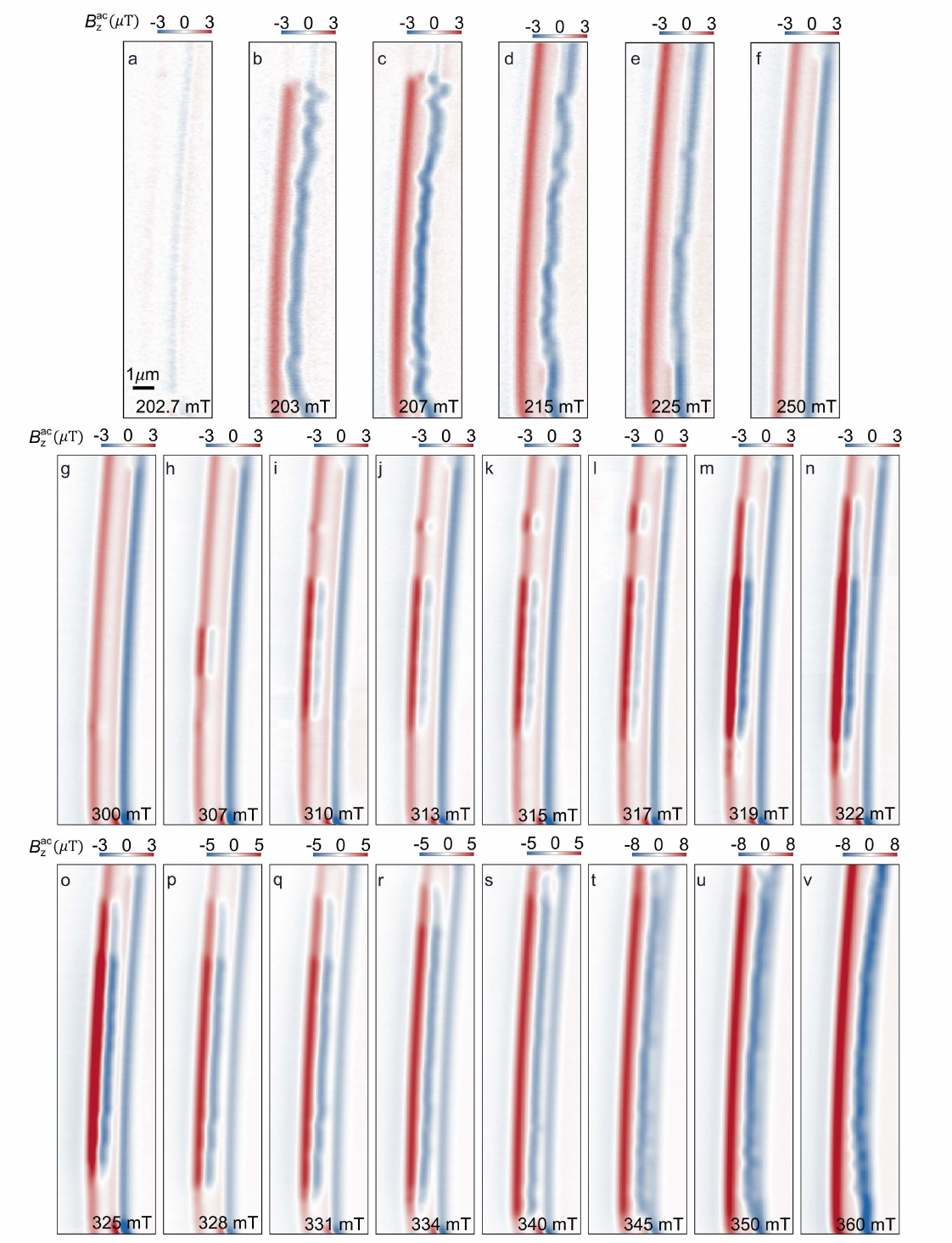}
\caption{\textbf{Stray-field images of the strained ribbon while sweeping \(H_x\) up}. $B_z^\text{ac}(x, y)$ with progressively increasing \(H_x\). (a) - (f) corresponds to the reorientation of the surface layers and (g) - (v) of the interior layers. }\label{figA4_strain_inplane_upBac}
\end{figure}

\begin{figure}[H]%
\centering
\includegraphics[width=\textwidth]{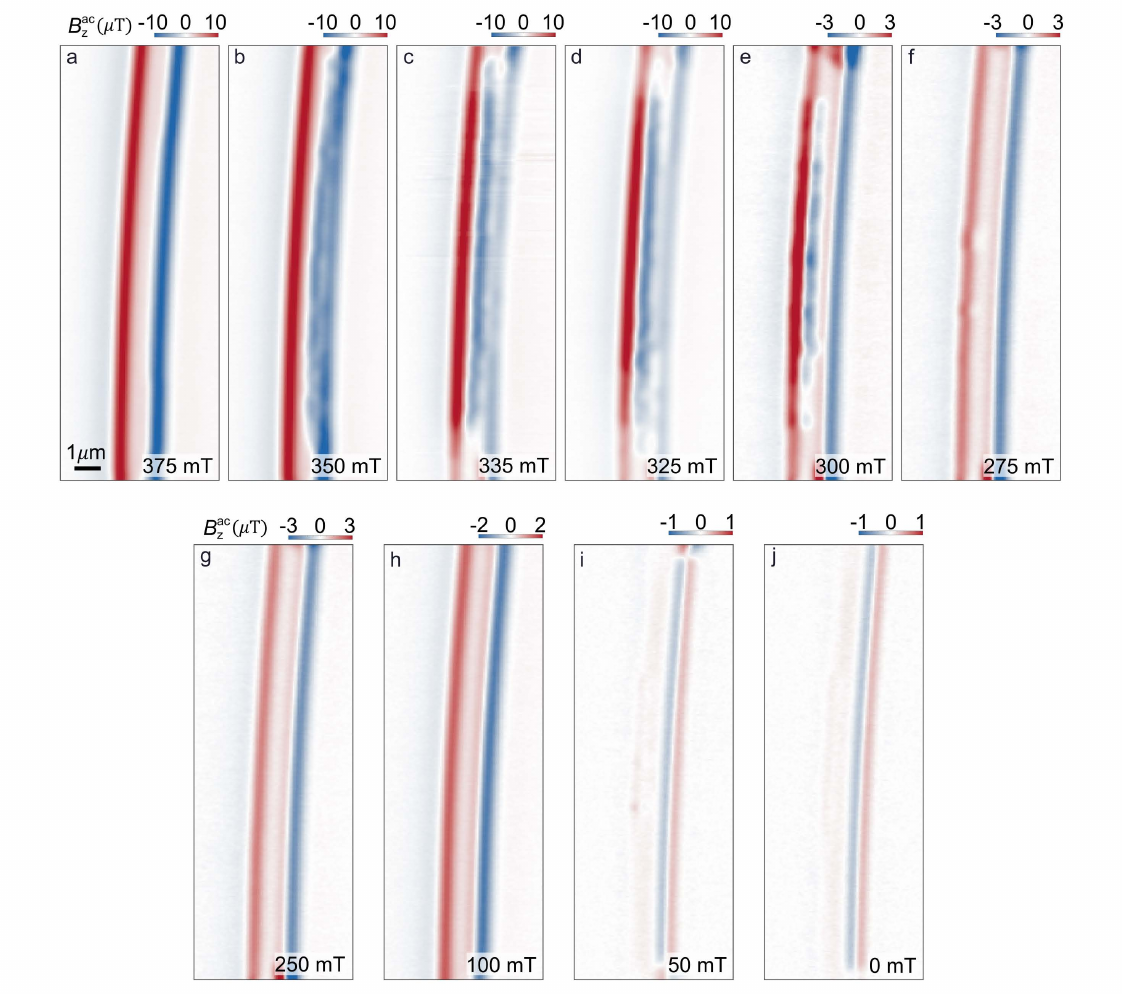}
\caption{\textbf{Stray-field images of the strained ribbon sweeping \(H_x\) down}. $B_z^\text{ac}(x, y)$ with progressively decreasing \(H_x\).}\label{figA4_strain_inplane_downBac}
\end{figure}

\begin{figure}[H]%
\centering
\includegraphics[width=\textwidth]{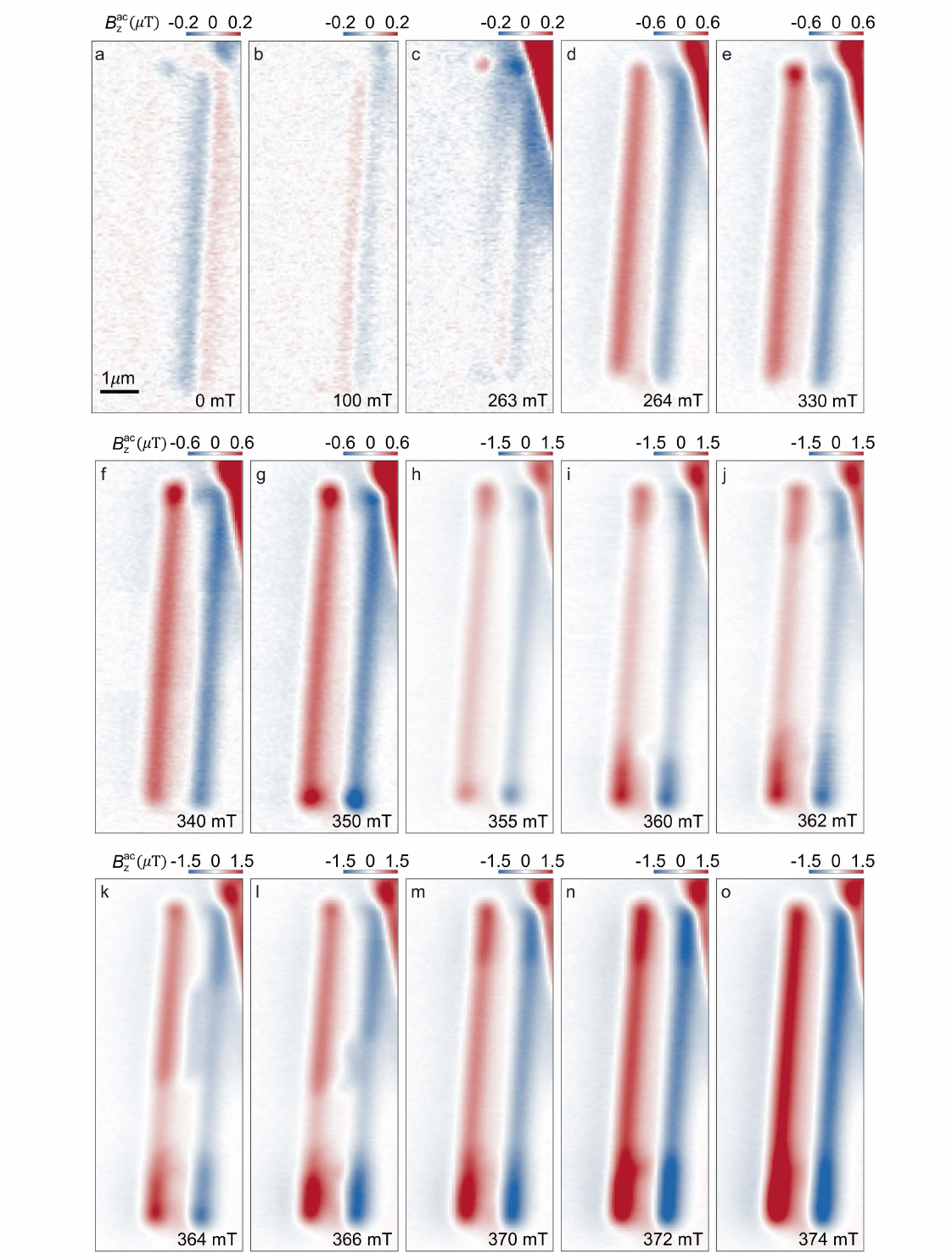}
\caption{\textbf{Stray-field images of the pristine ribbon sweeping \(H_x\) up}. $B_z^\text{ac}(x, y)$ with progressively increasing \(H_x\).}\label{figA4_pristine_inplane_upBac}
\end{figure}

\begin{figure}[H]%
\centering
\includegraphics[width=\textwidth]{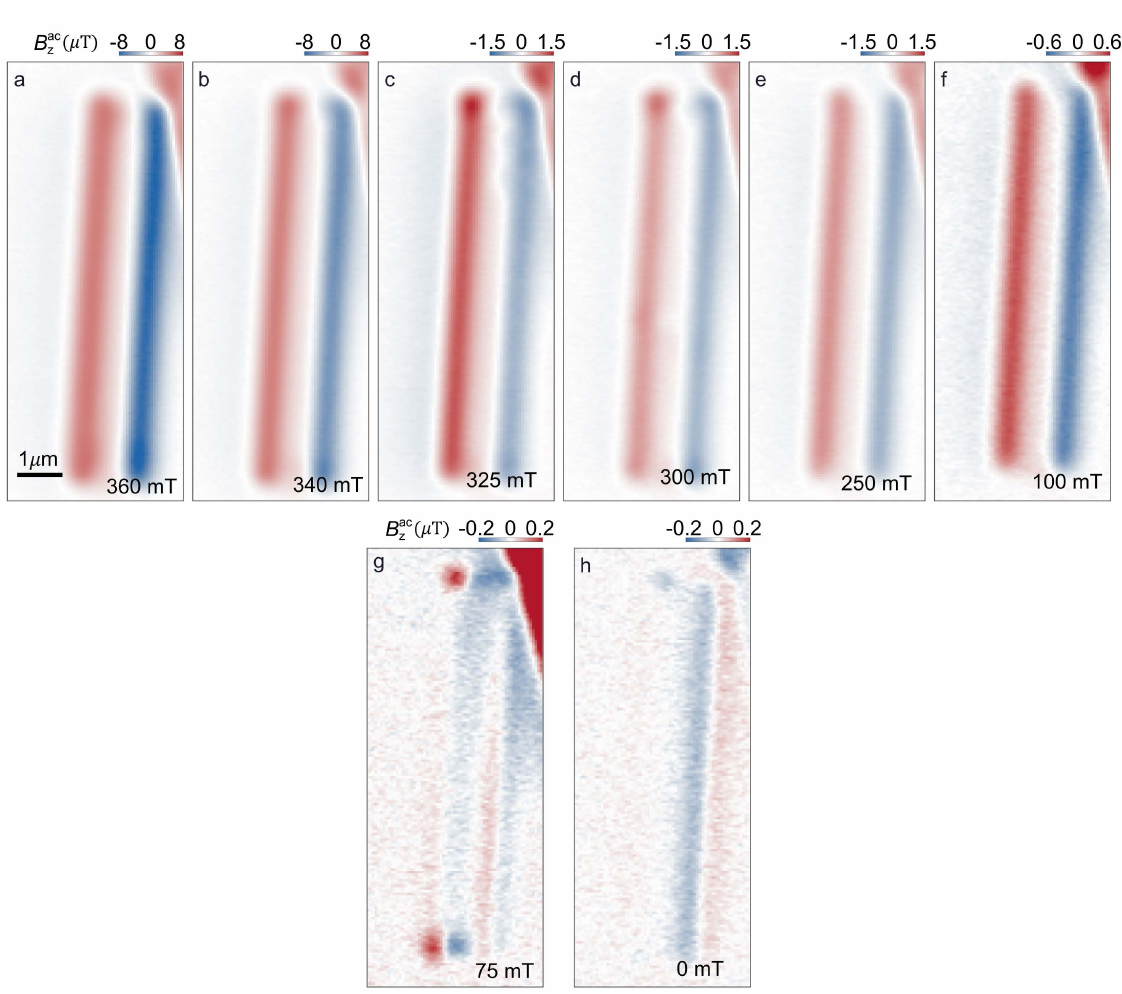}
\caption{\textbf{Stray-field images of the pristine ribbon sweeping \(H_x\) down}. $B_z^\text{ac}(x, y)$ with progressively decreasing \(H_x\).}\label{figA4_pristine_inplane_downBac}
\end{figure}

We image the local magnetic response of the CrSBr ribbons to a changing field $H_x$, applied nearly along the b-axis. 
In addition to the measurements shown in Figure~2 of the main text, Supplementary Figures~\ref{figA4_strain_inplane_upBac} and \ref{figA4_strain_inplane_downBac} show magnetic images of the strained ribbon measured as $H_x$ is swept both upwards and downwards. Similarly, Supplementary Figures~\ref{figA4_pristine_inplane_upBac} and \ref{figA4_pristine_inplane_downBac} show images of the pristine ribbon as $H_x$ is swept both upwards and downwards. We saturate both ribbons at $\mu_0H_x=-500$~mT before sweeping the field to zero and starting the upward measurement. Before the downward measurement, the ribbons were fully polarized in an applied field  $\mu_0H_x=600$~mT.
In all cases, we measure both $B_z$ and $B_z^\text{ac} \propto d B_z / d z$, however, we only plot $B_z^\text{ac}(x,y)$ maps, because they are more sensitive to small spatial features compared to $B_z(x,y)$.

\pagebreak

\section{Supplementary Note 2:\\ Magnetic hysteresis}\label{secA2}

A comparison between the magnetic hysteresis of the strained and pristine ribbons is shown in Supplementary Figure~\ref{figA4_hist}, where it is evident that the pristine ribbon (red and blue dashed lines) exhibits larger hysteresis compared to the strained ribbon (orange and light blue lines). For the pristine ribbon, we average  the majority of its area, excluding the top part of the ribbon to avoid the influence of the neighboring ribbon's stray field.  For the strained ribbon, we use the same area selected in Figure 4a.

As discussed in the main text, we infer that the transition occurring between 200 and 300~mT on the way up corresponds to the reorientation of a single layer. This conclusion is supported by the ratio of approximately 1:9 between the step in $|B_z|_\text{avg}$ associated with this transition and the value of $|B_z|_\text{avg}$ after the ribbons reach saturation (at around 400 mT), as shown in Supplementary Figure~\ref{figA4_hist}. 

\begin{figure}[H]%
\centering
\includegraphics[width=1\textwidth]{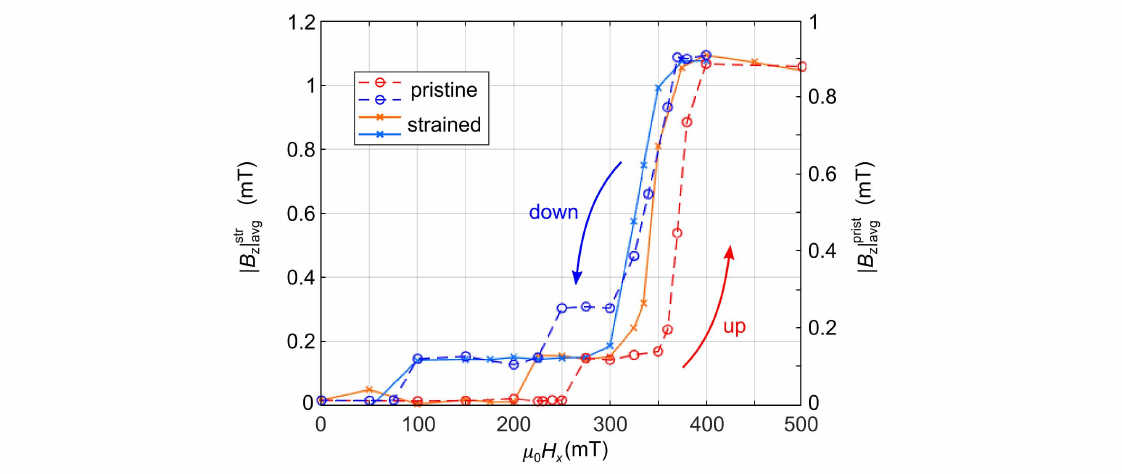}
\caption{\textbf{Magnetic Hysteresis comparison.} The vertical axes corresponds to $|B_z|$ averaged over areas of the pristine and strained ribbon and is plotted as a function of \(H_x\). Magnetic hysteresis of the pristine flake is shown in red and blue dashed lines for the upward and downward sweeps, respectively. The hysteresis of the strained flake is shown in orange and light blue lines for the upward and downward sweeps, respectively. }\label{figA4_hist}
\end{figure}

\pagebreak

\section{Supplementary Note 3:\\ Micromagnetic simulations}\label{secA3}

\begin{figure}[H]%
\centering
\includegraphics[width=1\textwidth]{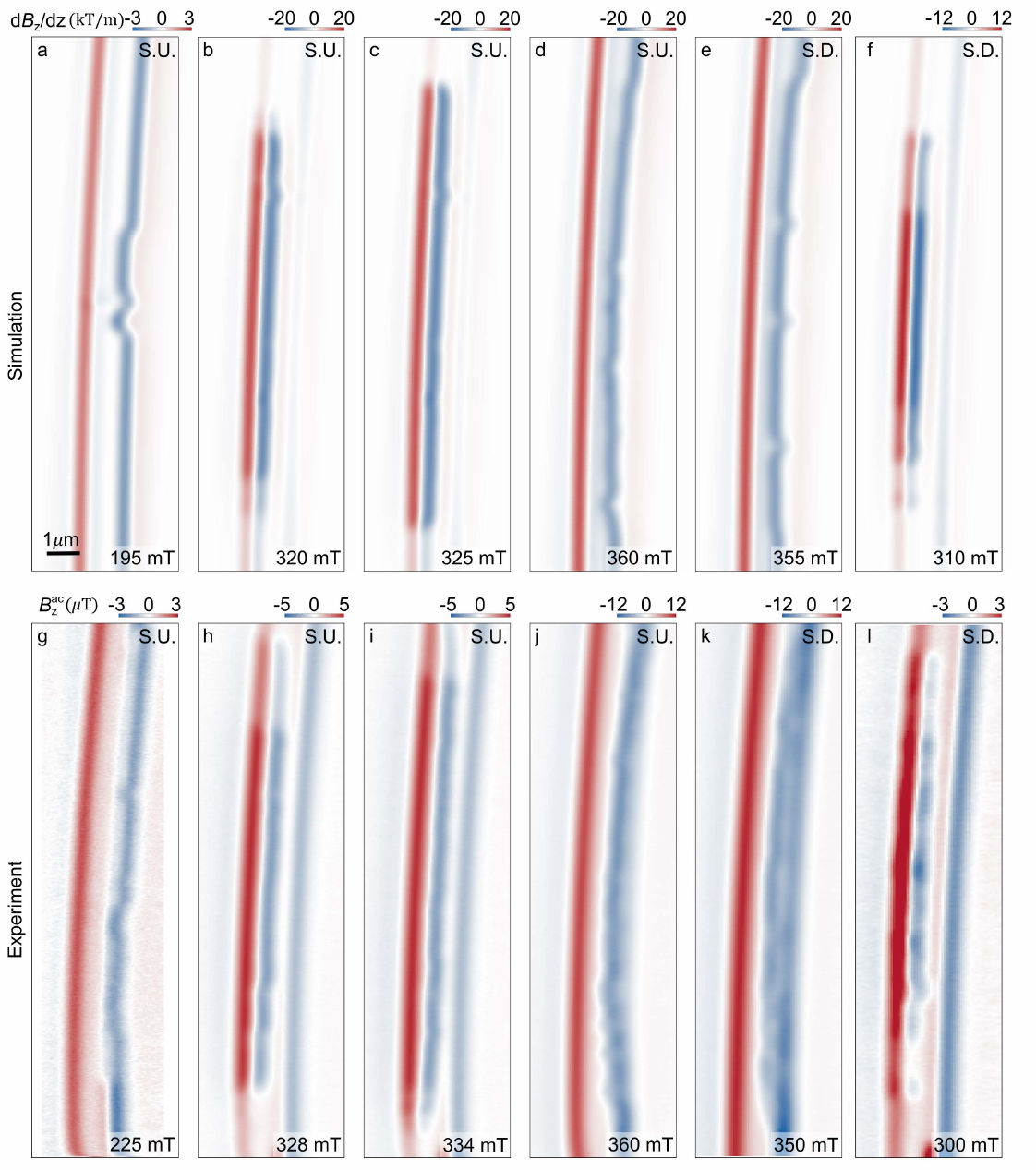}
\caption{\textbf{Comparison between simulated and measured magnetic domains. } (a) - (f) Representative simulated $\frac{\text{d}B_z(x, y)}{\text{d}z}$ at the applied fields $H_x$ indicated in the bottom right of the images and (g) - (l) corresponding measurements at similar fields. Whether the maps correspond to simulations or measurements done during a sweep up (S.U.) in $H_x$ or a sweep down (S.D.) is indicated in the top right of each image. }\label{S6}
\end{figure}

Micromagnetic simulations reproduce magnetic domains of the same shape and size as measured in experiment. This agreement suggests that the wavy features in the measured $B_z^\text{ac}(x,y)$ maps do not result from unintentional inhomogeneties or defects in the sample. Rather, they are a consequence of the strain gradients induced by bending the ribbon. Supplementary Figure~\ref{S6} shows a comparison between measured and simulated stray-field maps showing similar features at similar values of applied $H_x$.

\vspace{10mm}

%\pagebreak

%\bibliography{CrSBr}
%\addbibresource{CGT}